**SOLAR ABUNDANCE OF ELEMENTS FROM NEUTRON-CAPTURE CROSS SECTIONS.** O. Manuel[1], W. A. Myers[2], Y. Singh[3] and M. Pleess[4], [1]Nuclear Chemistry, University of Missouri, Rolla, MO 65401, USA, om@umr.edu, [2]Department of Chemical Engineering, University of Arkansas, Faretteville, AR 72701, USA, wmyers@engr.uark.edu, [3]Department of Computer Science, University of Missouri, Rolla, MO 65401, USA, ysxr6@umr.edu, [4]Kepler-Gymnasium, Tuebingen, Dettenhausen 72135, GERMANY, marcelpleess@web.de.

**Introduction:** Light elements, H and He, covering the solar surface are widely believed to be dominant inside the Sun [1]. Noting that gaseous envelopes may not reflect the Sun's internal composition, Harkins [2] used analyses on 443 meteorites to conclude that Fe, O, Ni and Si are more abundant. B2FH [3] explained element synthesis in stars; decades later Fowler said laboratory and theoretical calculations cannot explain why O/C = 2 [4]. Solar-wind-implanted elements in lunar soils showed a mass-dependent excess of lighter isotopes [5]. This empirical fractionation also yielded Fe, O, Ni and Si as the most abundant elements in the Sun [5, 6]. Neutron-capture cross-sections and the photospheric abundance [1] of s-process nuclides [3] offer another test for mass separation in the Sun.

**Composition of the Sun:** Below is a widely-quoted abundance pattern of elements in the Sun [1].

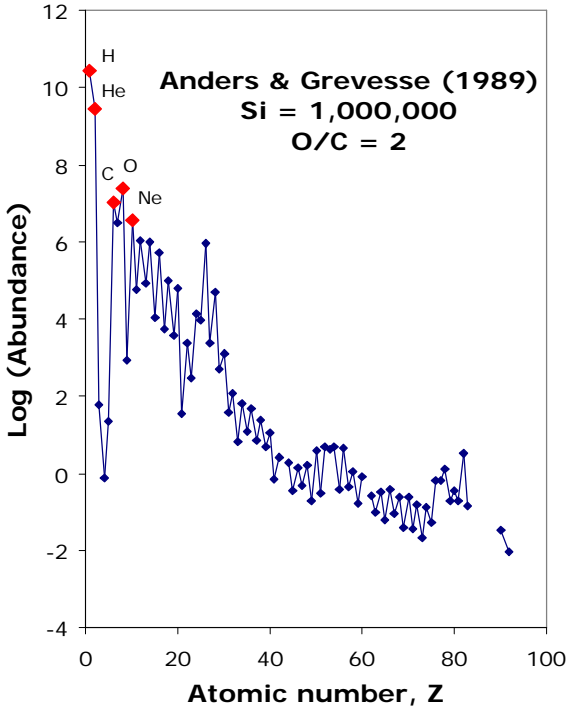

FIG 1. The abundance pattern of elements at the solar surface [1]. Large diamonds identify the more abundant elements.

Steady-flow abundances (N) and cross sections ($\sigma$) of nuclides made by the s-process are related [3]:

$$N_{(A-1)}\sigma_{(A-1)} = N_{(A)}\sigma_{(A)} = N_{(A+1)}\sigma_{(A+1)} \qquad (1)$$

Abundances of s-only isotopes of Sm [7] and Te [8] in the mass regions of 148-150 amu and 122-124 amu, respectively, confirm steady-flow. However, values of $N\sigma$ at the solar surface decrease by several orders of magnitude as A increases from 25 to 207 amu [3].

The exponential, least-squares line in FIG 2 shows the mass fractionation relationship in 72 s-products at the solar surface over the mass range of 25-207 amu.

$$\text{Log}(N\sigma) = -5.16 \log(A) + 12.5 \qquad (2)$$

The scatter is reduced, but the least-squares line defined by 20 "s-only" products from a more recent tabulation [9] over the 70-204 amu mass range has a similar slope. This is shown by Eq. 3 below the figure.

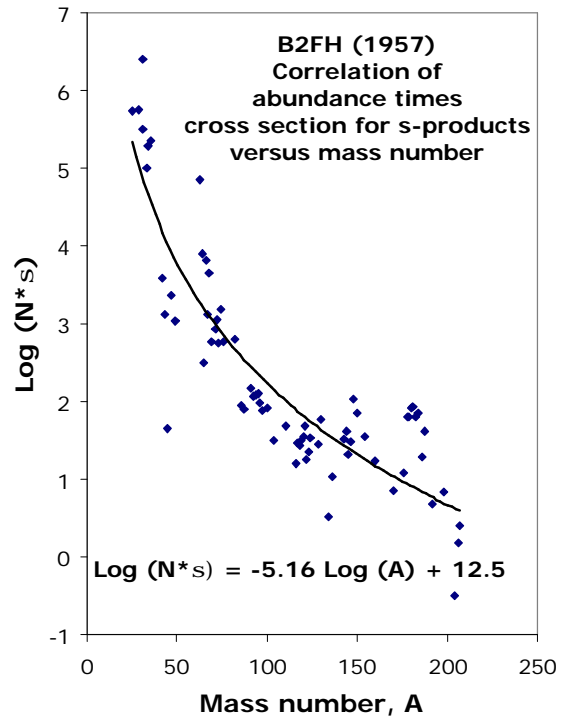

FIG 2. Values of $N\sigma$ for s-products at the solar surface decline exponentially with increasing mass number, A.

$$\text{Log}(N\sigma) = -5.14 \log(A) + 12.2 \qquad (3)$$

Thus the mass dependent fractionation relationship in FIG 2, eq. (2), and eq. (3) all show lighter mass (L) s-products enriched in the photosphere relative to heavier mass (H) ones by about a factor, $F$, where

$$F = (H/L)^{5.15} \quad (4)$$

On the other hand, light isotope enrichments in the solar wind [5] revealed an enrichment factor, $F$, of

$$F = (H/L)^{4.56} \quad (5)$$

FIG 3 is a comparison of these two empirical indications of mass separation that occurs inside the Sun.

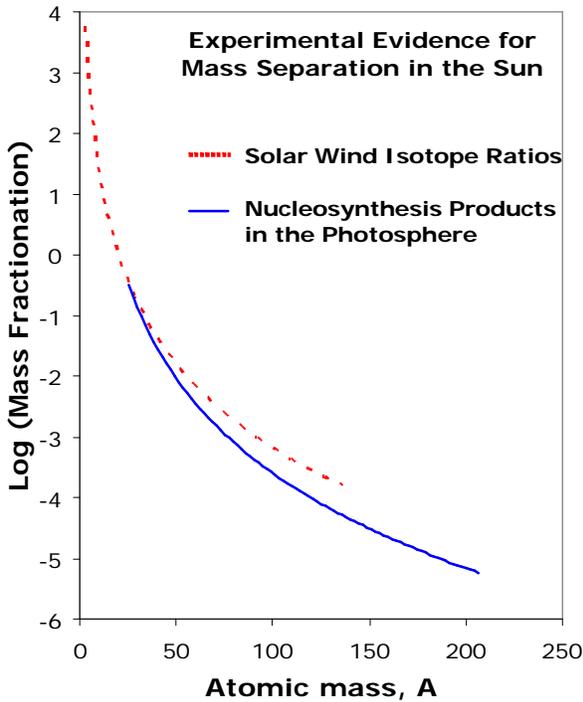

FIG 3. The dashed line, from 3 to 136 amu, shows the mass separation observed in solar wind isotopes [5]. The solid line, from 25 to 207 amu, shows the mass separation observed across s-products in the photosphere

**Conclusions:** Both empirical mass separation relationships indicate that Fe, O, Ni, and Si are abundant inside the Sun - the same elements that comprise most material in ordinary meteorites [2]. Fowler [4] identified the solar O/C ratio as one of two *"serious difficulties in the most basic concepts of nuclear astrophysics"* [4]. Since the O/C ratio is in the mass range of solar wind isotopes illustrated by the dashed line in FIG 3, that mass fractionation relationship is used to obtain the concluding elemental composition and the O/C ratio for the bulk Sun shown in FIG 4.

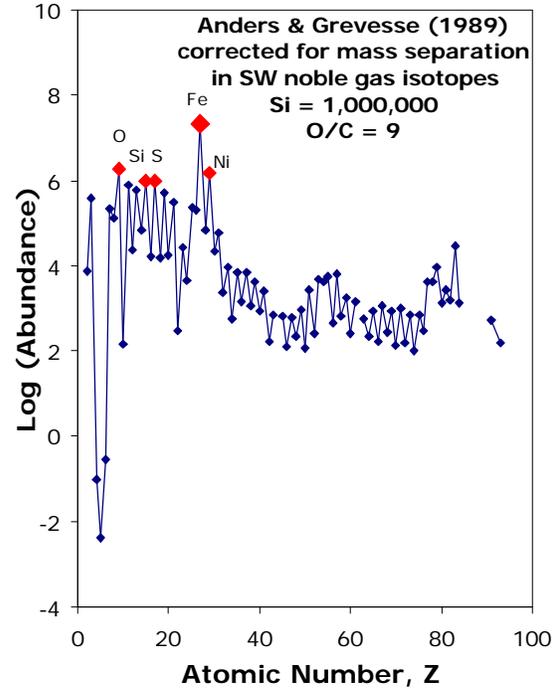

FIG 4. The composition of the Sun after correcting for the mass fractionation observed across isotopes in the solar wind [5]. O/C = 9 for the bulk Sun and its most abundant elements are Fe, O, Ni, Si and S.

The mass fractionation relationship defined by s-products in the photosphere (FIG 2) yields O/C = 10 for the bulk Sun. It also shows Fe, O, Ni and Si as abundant elements in the interior of the Sun. Our results may explain why heavy elements are abundant in impulsive solar flares [10] and resolve a long-standing difference between the solar photospheric value of O/C = 2 and calculated abundances of He-burning products [11].

**References**: [1] Anders E. and Grevesse N. (1989) *GCA, 53,* 197-214. [2] Harkins W. D. (1917) *J. Am. Chem. Soc., 39,* 856-879. [3] B2FH (1957) *Rev. Mod. Phys., 29,* 547-650. [4] Fowler W. A. in Forward to Rolfs C. E. and Rodney W. S. (1988) *Cauldrons in the Cosmos: Nuclear Astrophysics* (University of Chicago Press) p. xi. [5] Manuel O. and Hwaung G. (1983) *Meteoritics, 18,* 209–222. [6] Manuel O., Bolon C., Zhong, M. and Jangam P. (2002) *LPS XXXIII,* Abstract #1401. [7] Maclin R. L. et al. (1963) *Nature, 197,* 369-370. [8] Kaeppeler F. et al. (1982) *Ap. J., 257,* 821-846. [9] Seeger P. A. et al. (1964) *Ap. J. Suppl., 11,* 121-166. [10] Reames D. V. (2000) *Ap. J., 540,* L111-L114. [11] Deinzer W. and Salpeter (1964) *Ap. J., 140,* 499-509.

**Additional Information:** FCR, Inc., UMR and the University of Arkansas supported this research.